# Fast code CASCIE (Code for Accelerating Structures -Coupled Integral Equations). Test Results

*M.I. Ayzatsky*
*National Science Center Kharkov Institute of Physics and Technology (NSC KIPT), 610108, Kharkov, Ukraine*
*E-mail: mykola.aizatsky@gmail.com*

Results of testing the fast CASCIE (Code for Accelerating Structures - Coupled Integral Equations) code developed as an analytical-numerical tool for studying the properties of inhomogeneous structured waveguides are presented. We have used this code to numerically tune the resonator chain, which is very similar to the classic 3m SLAC accelerating structure. During testing, we discovered a problem that, in our opinion, was overlooked in the development of the accelerating sections, namely the appearance of a constant phase shift when determining the dimensions of the resonator radii from the characteristics of homogeneous structures. This issue is particularly relevant in the development of short accelerating structures.

## 1. INTRODUCTION

Among the different waveguides, there is the class of structured waveguides - waveguides that consist of similar (but not always identical) cells. Structured waveguides can be closed [[1,2,3,4,5,6,7,8,9,10,11,12,13]] (wave propagation in the transverse direction is limited by the walls) and open (evanescence in the transverse direction is determined by the electrodynamic properties of the medium) [14,15]. There are powerful codes that can be used to calculation RF characteristic of structured waveguides, but it is difficult to use them for investigation properties and making preliminary design. For these purposes the fast and versatile models are needed.

Structured waveguides based on coupled resonators have propagation characteristics similar (or identical for the homogeneous case) to one-dimensional Floquet-Bloch band structures. Despite their wide application and long period of study, the description of their properties, especially inhomogeneous, is far from complete. The same applies to the mathematical models that are used to study. The main difficulties in the study such waveguides are based on the fact that they are two-dimensional and there are no appropriate sets of one-dimensional functions that can be used for representing the fields. This manifests itself most clearly when calculating the characteristics of inhomogeneous resonator chains.

Taking into account this circumstance, it was proposed to use difference equations to describe the inhomogeneous resonator chains. The first attempt was made on the base of the coupled cavities model that was developed with using many eigen modes and rigorous calculation of coupling coefficients [16]. Obtained difference equations that connect the values of electric field in different points of resonators correctly describe the main waves, but also contain different spurious oscillations [17].

To explore other possibilities of using difference equations and approximate methods, we have proposed a simple but rigorous model of chain of cylindrical resonators coupled through openings in the very thing walls [18]. This model is based on the method of Coupled Integral Equations (CIE) (see, for example, [19,20]). Using the theory of solving matrix equations (see [21,22] and sited there literature) and the decomposition method [23], we obtained new matrix difference equations, on the basis of which various approximate approaches, including the WKB approach, were developed. In this model for correct representing the fields it was enough to use two functions with singularities at the rim of openings for calculating the electric fields on the dividing surfaces.

It is worth to note that the unknowns in the matrix difference equations are vectors which components are the moments of electric fields on the surfaces that divide the chain resonators. Determining these moments gives possibility to calculate electromagnetic fields in any point of resonator. Therefore, proposed equations are not direct equations for the electric field.

For the case of finite thickness walls between resonators there are no simple analytical functions that describe correctly the singularities of electric fields near the edge of openings and give simple integrals [24]. The use of basis functions without singularities leads to the need to use a large number of terms in the series. For long chains, this becomes one of the major problems [25,26].

To eliminate this shortcoming, we have proposed some modifications to the standard approach [27]. In earlier developed schemes the coupled integral equations were derived for the unknown electrical fields at interfaces that divide the adjacent volumes. In addition to the standard division of the structured waveguide by interfaces between the adjacent cells, we propose to introduce new interfaces in places where electric field has the simplest transverse structure. Moreover, the system of coupled integral equations is formulated for longitudinal electrical fields in contrast to the standard approach where the transverse electrical fields are unknowns. The final vector equations contain expansion coefficients of the longitudinal electric field at these additional interfaces. This modification makes it possible to deal with a physical quantity (longitudinal electric field), which plays an important role in tuning accelerator structures and particle acceleration, and to obtain approximate equations for the case of a slow change in the waveguide parameters. As calculation results have shown, the number of the necessary coefficients decreased significantly [27].

Tuning of TW normal conducting sections is an important step in the process of its manufacturing. There is no general approach on how to carry out such tuning. It still is more art than science. Some results have been obtained for constant gradient sections [28,29,30].

Moreover, there is no answer to the question of what laws of group velocity (opening radii) variation can be implemented under the specific requirements of the accelerating process. For getting new results in this problem without the use of complex numerical codes and big computers, the simple and fast code is needed.

In this paper we present the results of testing of a fast code CASCIE (Code for Accelerating Structures - Coupled Integral Equations) which was developed on the base of proposed modified method of Coupled Integral Equations [27].

## 2. MODEL EQUATIONS

Let us consider a chain consisting of segments of cylindrical waveguides of different radii. On each side, the chain is connected to a semi-infinite cylindrical waveguide. All waveguides share a common z-axis. We will consider only axially symmetric fields with $E_z, E_r, H_\varphi$ components (TM). Time dependence is $\exp(-i\omega t)$. We also suppose that in the left semi-infinite waveguide the $TM_{0,1}$ eigen wave propagates towards the considered chain ( $E_z = J_0(\lambda_{01} r / b_w) \exp(i\, h_w z)$ ).

Even segments will be numbered by the index $k$ ( $1 \leq k \leq N_{REZ}$ ) and have radii $b_k$ and lengths $d_k$, Odd segments will be numbered by the index $k'$ ( $1' \leq k' \leq (N_{REZ} + 1)'$ ) and have radii $a_{k'}$ and lengths $d_{k'}$ A segment placed to the left of waveguide with an index $k$, will have $k' = k$.

In this paper we will consider the case $a_{k'} < b_k$ and the chain can be considered as the chain of resonators coupling through cylindrical openings in the walls of finite thickness. Below, segments of larger radius will be called resonators.

The longitudinal component of electric field at the cross section that locates at the middle of even segments (resonators) we expand in terms of a set of basis functions $\varphi_s^{(z,k)}(r/b_k)$

$$E_z^{(k)}(r, d_k/2) = \sum_s Q_s^{(k)} \varphi_s^{(z,k)}(r/b_k). \tag{1}$$

As the longitudinal component of electric field has not singularities nearby this cross section and $E_z^{(k)}(b_k, d_k/2) = 0$, the choice can be

$$\varphi_s^{(z,k)}(r/b_k) = J_0(\lambda_s r / b_k), \tag{2}$$

where $J_0(\lambda_s) = 0$.

Using the Coupled Integral Equations Method, such equations can be obtained [27]

$$\begin{aligned}
&T^{(Q_1)} Q^{(1)} + T^{(Q_2)} Q^{(2)} = Z^{Q(1)}, \\
&T^{(k)} Q^{(k)} = T^{+(k)} Q^{(k+1)} + T^{-(k)} Q^{(k-1)} + Z^{Q_k},\ k=2,...,N_{REZ}-1, \\
&T^{(Q_{NREZ}-1)} Q^{(N_{REZ}-1)} + T^{(Q_{NREZ})} Q^{(N_{REZ})} = Z^{Q_{NREZ}},
\end{aligned} \tag{3}$$

where $T^{(k)}$ are the matrices (see [27]) and $Q^{(k)} = \left(Q_1^{(k)}, Q_2^{(k)}, Q_3^{(k)}, ....\right)^T$ are the vectors, $Z^{Q_k}$ - vectors which are defined by the incident wave and electron currents. We cannot solve such equations since $T^{(k)}$ and $Q^{(k)}$ have infinitive dimensions. As usual, we truncate these objects to some finite size: $T^{(k)} \Rightarrow T^{(k)} \in C^{(N_z, N_z)}, Q^{(k)} \Rightarrow Q^{(k)} \in C^{(N_z)}$ The question is then how do we determine the $N_z$ so that the results are good to some approximation. It was shown that at frequency $\omega_0 = 2\pi\, 2.856$ GHz and $a_{k'} < 1.5$ cm the reasonable values of $N_z$ is 4÷5 (see below).

## 2. SIMULATION RESULTS

First of all, we introduce two functions that relate the geometric dimensions and parameters of the waves in the main passband of a homogeneous periodic chain of lossless waveguides

$$a = f_a(v_g, \varphi, \omega, d_a, d_b), \tag{4}$$

$$b = f_b(a), \tag{5}$$

where $a$ and $b$ ( $a < b$ ) are the radii of the waveguide segments constituting one period, $d_a, d_b$ are the lengths of these waveguides, $v_g$ and $\varphi$ are the group velocity and the phase shift per period.

Numerical calculations[1] (see Figure 1 and Figure 2) for the case when $d_a = 0.5842$ cm, $d_b = 2.9147$ cm ($D = d_a + d_b = \lambda_0/3$), $\omega_0 = 2\pi \cdot 2.856$ GHz, and $\varphi = \dfrac{2\pi}{3}$ show that for $0.005 < \beta_g = v_g/c < 0.02$ we can use such dependencies ($[a] = cm, [b] = cm$)

$$\beta_g = 0.0364a^2 - 0.0491a + 0.0184,$$
$$a = 0.6745 + \sqrt{-0.0506 + 27.4745\beta_g} \tag{6}$$
$$b = 0.1576a^2 - 0.1476a + 4.0652. \tag{7}$$

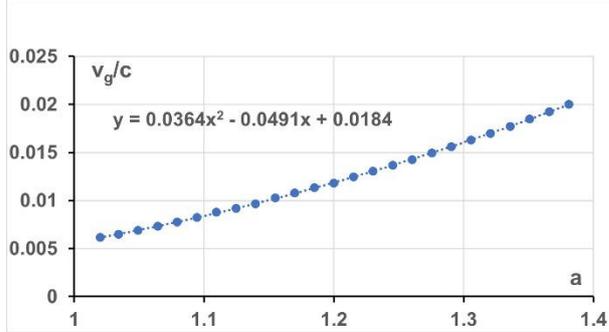 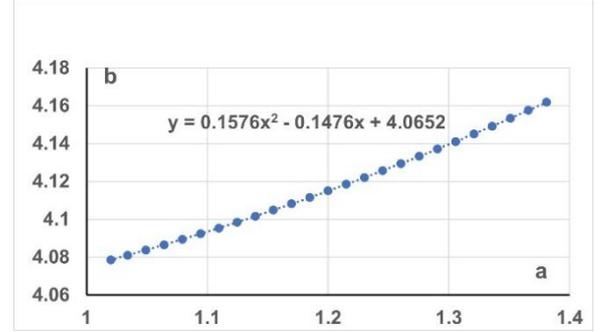

Figure 1 Group velocity as function of radius $a$ for the periodic chain.

Figure 2 Radius $b$ as function of radius $a$ for the periodic chain.

We will consider a linear law of change $v_g$

$$v_{g,k'} = \frac{v_{g,1} + v_{g,2}}{2} + \frac{v_{g,1} - v_{g,2}}{2} \frac{(k' - k_0)}{(3 - k_0)}, \tag{8}$$

where $k' = 3,...,N_{REZ} - 1$, $k_0 = N_{REZ}/2 + 1$.

The law of change in the radius of even sections (resonators) can be arbitrary, but we choose it according to the methodology used in the development of accelerating sections - the radii of resonators correspond to the "homogeneous" case

$$b_k = f_b(a_k), \; k = 3,...,N_{REZ} - 1. \tag{9}$$

Under such choice, one can expect that the phase shifts between the resonators will be about $2\pi/3$. Below we will consider the case $\beta_{g,1} = 0.02$, $\beta_{g,2} = 0.0062$.

In the waveguides electromagnetic energy dissipates in walls. Taking into account dissipation is not simple task, especially for semi-analytical approaches. To simplify the task, we filled all waveguide segments with dielectric with complex permittivity $\varepsilon = \varepsilon' + i\varepsilon''$, $\varepsilon'' > 0$. In this case we can use the same sets of functions as in a lossless task. Usually, we took $\varepsilon' = 1$ and $\varepsilon'' = const$ for all values of radii $a_{k'}$ and $b_k$. The value of $\varepsilon''$ was chosen from the condition $\alpha(\varepsilon'', \beta_{g,2}) \approx \alpha_{SLAC-END}$, where $\alpha$ is the field attenuation per unit length, $\alpha_{SLAC-END}$ is the attenuation for the last cell of the SLAC 3 meter structure ($\beta_g = 0.0061$, $Q = 13220$ [28]). Calculation of the dependence of $\alpha$ on $\beta_g$ (see Figure 3 and Figure 4, $\varepsilon'' = 7.3 \cdot 10^{-5}$) shows that such choice gives good coincidence the value of $\alpha$ at $\beta_{g,1} = 0.02$ and the SLAC section entrance value ($\beta_g = 0.0204$, $Q = 14170$). Moreover, we obtain the correct dependence of $\alpha$ on $\beta_g$ ($\alpha \sim 1/\beta_g$, see Figure 4).

---

[1] We used the EVCCG program from IMSL MATH Fortran Library to find the eigen values of the matrix that describes a homogeneous chain (see [27])

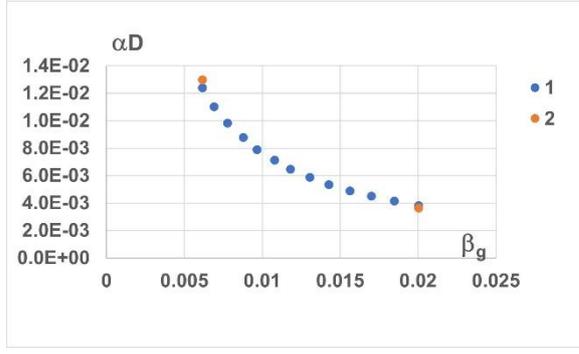 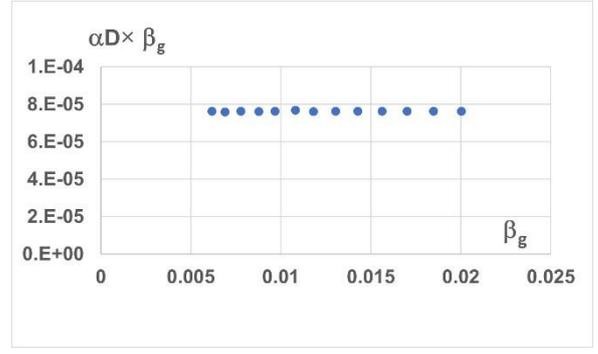

Figure 3 Dependence of $\alpha D$ on $\beta_g$, 1 - simulation results, 2-values for SLAC 3-meter structure [28].

Figure 4 Dependence of $\alpha D \beta_g$ on $\beta_g$.

We will consider the case when the radii of two semi-infinite waveguides are the same and that only the dominant mode $TM_{01}$ can propagate, while all higher-order modes are evanescent.

In this paper we present the results of studying the traveling wave regime. To realized it, the geometric dimensions of the first two segments of the chain were chosen from the condition of the absence of reflections from interface between semi-infinitive periodic chain of waveguides with $a_{k'} = const = f_a\left(v_{g,1}, \varphi = \frac{2\pi}{3}\right)$, $k' = -\infty,...,0$ and $b_k = const = f_b(a_k)$, $k = -\infty,...,1$ and the cylindrical waveguide with radius $b_w$. The geometric dimensions of the last two segments were chosen from the similar condition, but with $v_g = v_{g,2}$ ($v_{g,2} < v_{g,1}$). It is worth noting that input and output couplers must be tuned slightly differently [31,32]. We did not take into account this difference.

It should be noted that losses have some effect on the dimensions of the couplers. If we conduct numerical tuning of couplers with $\varepsilon'' = 0$ and $\varepsilon'' \neq 0$, we obtain notable differences (see Table 1), that can give additional reflections (see Figure 5 and Figure 6, $R$ - reflection coefficient - amplitude of the reflected wave in the left homogeneous waveguide)

| Table 1 | | |
|---|---|---|
| | $\varepsilon'' = 0$ | $\varepsilon'' = 7.310^{-5}$ |
| $a_{1'}$, cm | 1.9589 | 1.9575 |
| $b_1$, cm | 4.2333 | 4.2331 |
| $a_{(NREZ+1)'}$, cm | 1.6390 | 1.6421 |
| $b_{NREZ}$, cm | 4.1441 | 4.1445 |

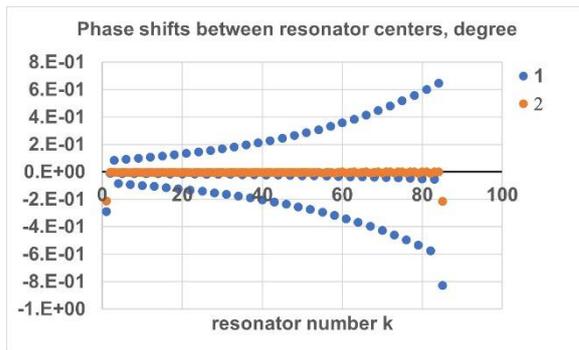 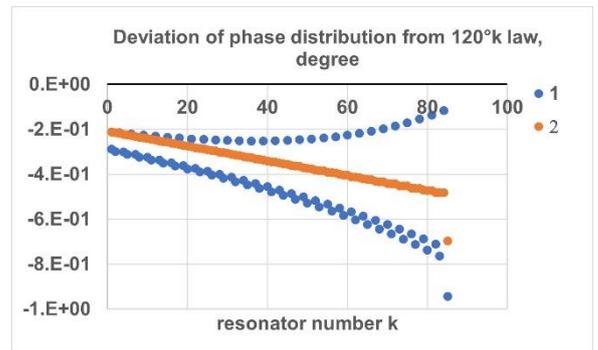

Figure 5 Phase shifts between centers of neighboring resonators in a homogeneous chain with $\beta_g$ =0.0061 (86 resonator chain); 1- the couplers were tuned with $\varepsilon'' = 0$, $R$ =3.3E-2, 2- the couplers were tuned with $\varepsilon'' = 7.310^{-5}$, $R$ =1.5E-2.

Figure 6 Deviation of phase distribution in the centers of resonators from $120°k$ law in homogeneous chain with $\beta_g$ =0.0061 (86 resonator chain); 1- the couplers were tuned with $\varepsilon'' = 0$, 2- the couplers were tuned with $\varepsilon'' = 7.310^{-5}$.

First of all, it is necessary to choose the number $N_z$ of the basis functions $\varphi_s^{(z,k)}(r/b_k) = J_0(\lambda_s r/b_k)$, $s = 1,...,N_z$, with which we can achieve desirable accuracy of field simulation (less than 1%). In the case of homogeneous chain, it was shown that simulation with $N_z$ =4 gives good accuracy both in calculating the longitudinal electric field and in calculating the dispersion characteristics [27]. To analyze the accuracy in the case of an inhomogeneous chain, we used two characteristics: the amplitude error $\Delta_E^{(k)} = \left( \left| E_z^{(k)}(N_{z,1}) \right| - \left| E_z^{(k)}(N_{z,2}) \right| \right) / \left| E_z^{(k)}(N_{z,1}) \right|$ and the phase error $D\varphi^{(k)} = \varphi^{(k)}(N_{z,1}) - \varphi^{(k)}(N_{z,2})$, where $N_{z,1}, N_{z,2}$ are the numbers of basis functions. Results of simulations, which are presented in Figure 7 and Figure 8, show that simulation with $N_z$=4 also gives good accuracy for inhomogeneous chain. At this value of $N_z$ the relative error in calculating the components of the electric field is less than a few units out of $10^{-3}$.

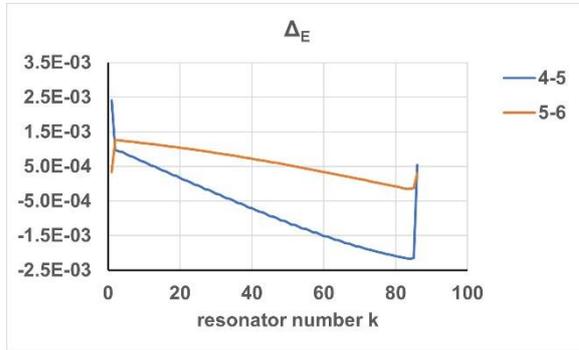
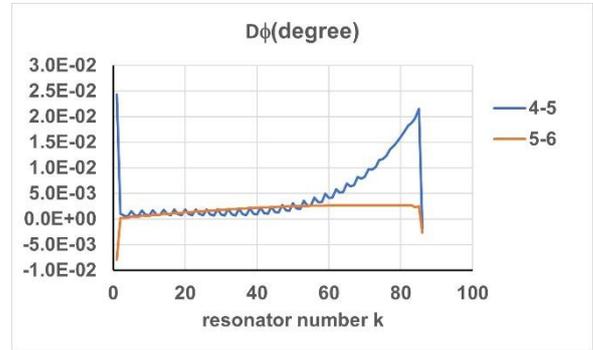

Figure 7 Amplitude error: 1 - $N_{z,1} = 4$, $N_{z,2} = 5$; 2 - $N_{z,1} = 5$, $N_{z,2} = 6$.

Figure 8 Phase error: 1 - $N_{z,1} = 4$, $N_{z,2} = 5$; 2 - $N_{z,1} = 5$, $N_{z,2} = 6$.

Consider the chain that consists of 86 pairs of waveguide segments (number of resonators $N_{REZ} = 86$). As $\beta_g$ decreases linearly, this chain is very similar to the classic SLAC section [28]. Initial variations of geometric parameters (in "homogeneous" approach) are presented in Figure 9 and Figure 10. The calculated values of the longitudinal electric field at the centers of the resonators are presented in Figure 11 and Figure 12. From the phase distribution (Figure 12, Figure 13, Figure 14) we can see that under the "homogeneous" approach in the inhomogeneous chain there is additional phase shift of the same sign between resonators (linear increase of phase deviation, see Figure 14). You can find some analysis of this phenomenon in Appendix 1.

We can cancel[2] this shift by introducing smooth linear decrease in the radii of the resonators[3]

$$b^{(k)} = 0.1576 a^{(k)2} - 0.1476 a^{(k)} + 4.0652 - \eta(1 + \gamma(k-4)), \ k = 4,...,83 . \quad (10)$$

By selecting the optimal values of $\eta$ and $\gamma$ it is possible to cancel the growth of the phase deviation (see Figure 13, Figure 14, Figure 16, Figure 17). After such a "correction", we have obtained a deviation of the phase distributions from the $2\pi/3$ law of less than ± 5 degrees. It must be noted that, until now, we have used the "monotonic" geometric $b$ distribution, with the exception of one jump between the 83rd and 84th resonators (see maximum value of $k$ in the expression (10) ). It can be seen from Figure 11 that this jump gives an almost biperiodic change in the amplitude. Therefore, additional tuning is required.

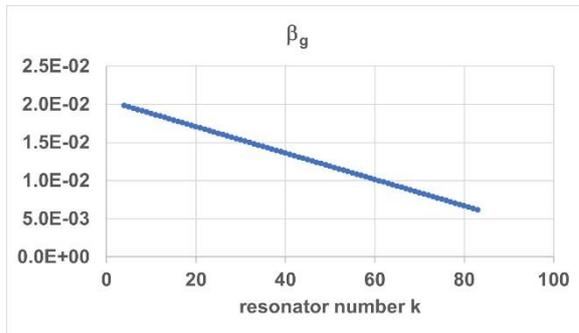
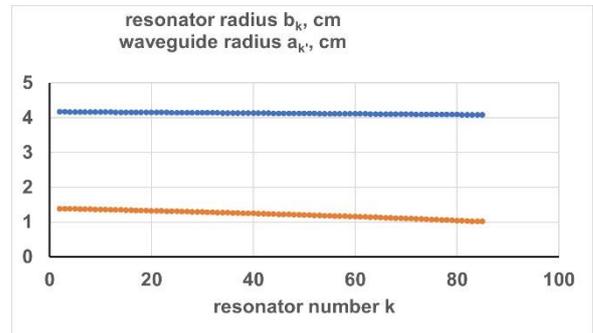

Figure 9 Specified values for group velocity.

Figure 10 Calculated values of segment radii.

---

[2] It can be also canceled by shifting the working frequency by 400 kHz.
[3] We did not resize the two resonators after the input coupler and the two resonators before the output coupler to avoid reflections from the couplers.

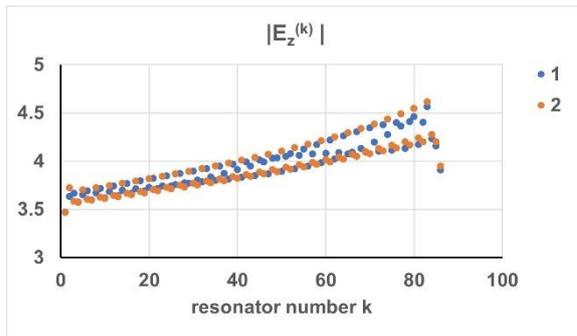
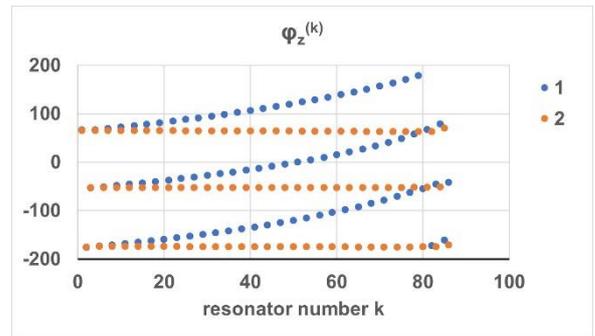

Figure 11 Amplitude of longitudinal electric field at the centers of resonators: 1 – the radii of resonators correspond to the "homogeneous" case (7), $R$ =2.7E-2, 2 - the radii of resonators correspond to the "corrected" case (10), $R$ =3.3E-2.

Figure 12 Phase distribution of longitudinal electric field at the centers of resonators: 1 – the radii of resonators correspond to the "homogeneous" case (7), 2 - the radii of resonators correspond to the "corrected" case (10).

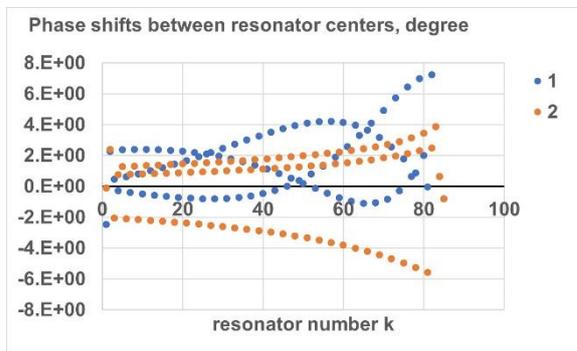
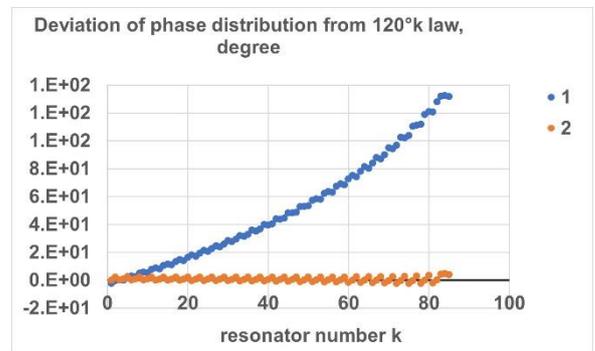

Figure 13 Phase shifts (relative $2\pi/3$) between the neighboring resonators: 1 – the radii of resonators correspond to the "homogeneous" case (7), 2 - the radii of resonators correspond to the "corrected" case (10).

Figure 14 Deviation of phase distributions from $2\pi k/3$ law 1 – the radii of resonators correspond to the "homogeneous" case (7), 2 - the radii of resonators correspond to the "corrected" case (10).

For today, there is no the general tuning procedure that can lead to a desired field distribution. It is a consequence of the fact that there is no a comprehensive understanding of the relationship between geometric dimensions $a^{(k)}$, $b^{(k)}$ and electric field distribution in inhomogeneous case.

There are two most using tuning methods: phase Ph-method and S-method. In the first method the phase shifts between resonators are tuned to the desired values by slightly changing of the cavity radii (see, for example, [28]). In the second method the combinations of field meanings in some points of several cells are reduced to the desired values by the same actions [29]. For more details of the S method and its restriction, see [30].

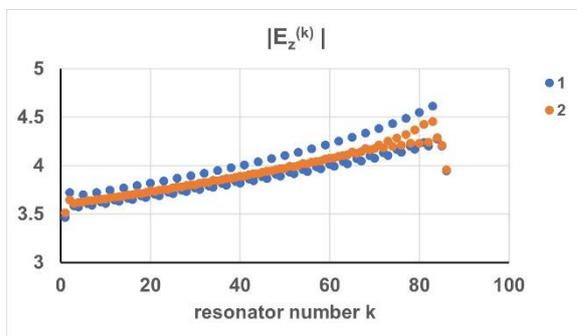
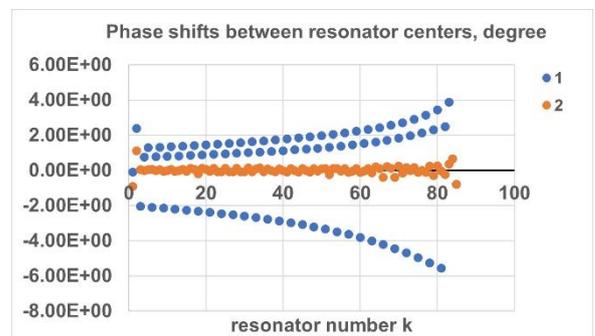

Figure 15 Amplitude of longitudinal electric field at the centers of resonators: 1 - the radii of resonators correspond to the "corrected" case (10) $R$ =3.3E-2; 2 – after tuning procedure, $R$ =1.4E-2.

Figure 16 Phase shifts (relative $2\pi/3$) between the neighboring resonators:1 - the radii of resonators correspond to the "corrected" case (10); 2 –after tuning procedure.

We used the modified phase Ph-method – we have corrected the deviation of phase distributions from the $2\pi k/3$ law by successively changing the radii of the resonators (from the 84th resonator to the 5th resonator). This tuning procedure cannot guarantee that the cavity radius distribution will be smooth. Results of such tuning are presented in Figure 15 - Figure 18. From Figure 15 it follows that after tuning a small amplitude oscillation occurs at the end of the section.

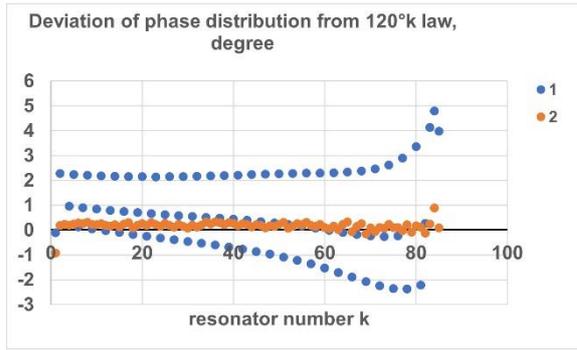

Figure 17 Deviation of phase distributions from $2\pi k/3$ law): 1 - the radii of resonators correspond to the "corrected" case (10); 2 –after tuning procedure.

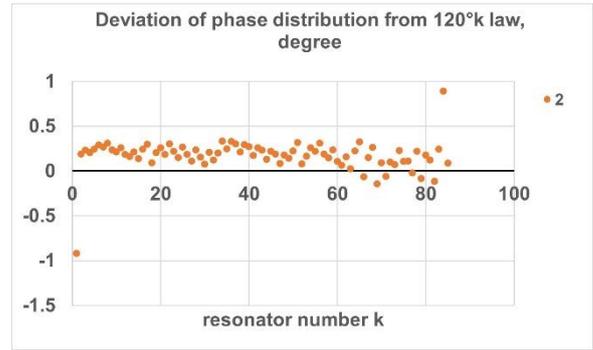

Figure 18 Deviation of phase distributions from $2\pi k/3$ law after tuning procedure .

If we apply the tuning procedure without preliminary correction (10), we get a good phase distribution (Figure 20), but amplitude oscillation will occur throughout the entire chain (see Figure 19).

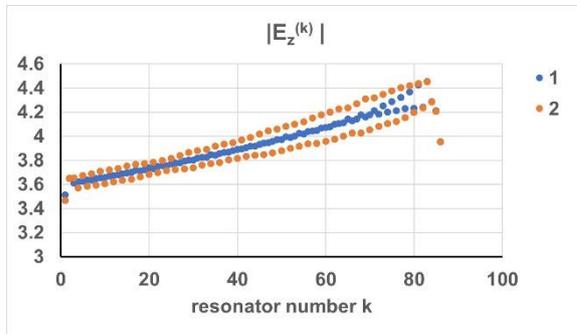

Figure 19 Amplitude of longitudinal electric field at the centers of resonators:1 - after tuning procedure in the case when the initial radii of resonators correspond to the "corrected" distribution (10), 2 - after tuning procedure in the case when the initial radii of resonators correspond to the "homogeneous" distribution (7).

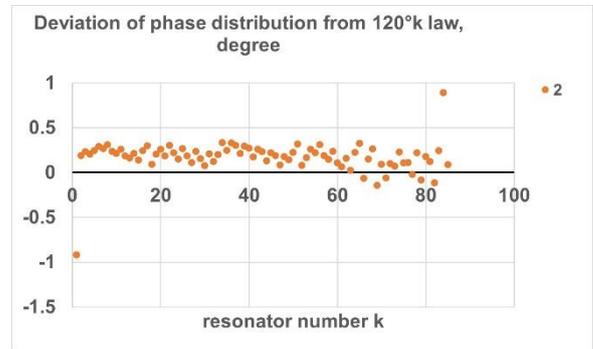

Figure 20 Deviation of phase distributions from $2\pi k/3$ law after tuning procedure in the case when the initial radii of resonators correspond to the "homogeneous" distribution (7).

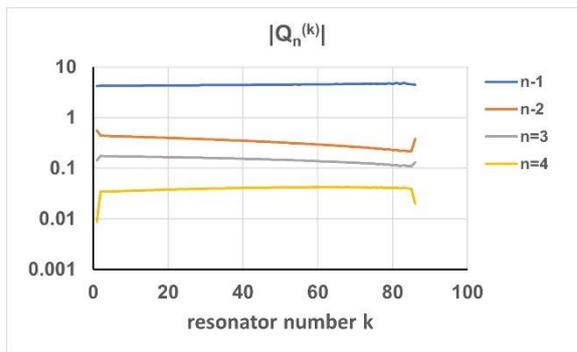

Figure 21 Expansion coefficients $Q_n^{(k)}$ of the longitudinal component of electric field after tuning for different resonators.

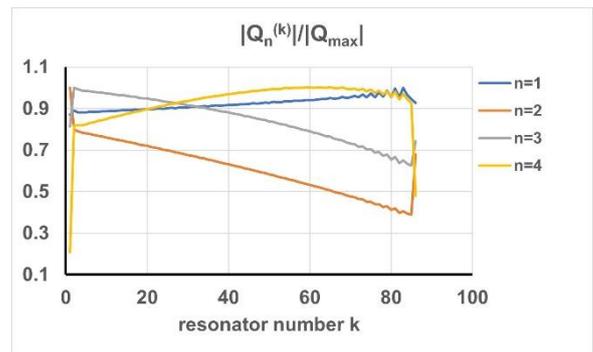

Figure 22 Normalized expansion coefficients $Q_n^{(k)}$ of the longitudinal component of electric field after tuning for different resonators.

One of the most important issues in the theory of waveguides is the possibility of using one variable (complex amplitude) for the analytical description of electromagnetic fields in the inhomogeneous structured waveguides.

It was shown that in matrix approach such possibility exists for homogeneous structured waveguides [27]. For inhomogeneous waveguides it could be possible if the amplitudes $Q_s^{(k)}$ in the expansion (1) could be presented as

$$Q_s^{(k)} = \bar{Q}^{(k)} q_s .  \qquad (11)$$

Then for $r = 0$ we would have

$$E_z^{(k)}(r=0, d_k/2) = \bar{Q}^{(k)} \sum_s q_s. \tag{12}$$

Therefore, if $Q_s^{(k)}$ as a function of $k$ is the same for all $s$ (up to the constant multiplier $q_s$), then electromagnetic fields can be described by a single variable. The simulation results presented in Figure 21 and Figure 22 show that representation (11) is incorrect. Therefore, there is a question about the accuracy of models that use one variable for the analytical description of electromagnetic fields in inhomogeneous structured waveguides (see, for example, [33,34,35,36,37,38,39,40]).

## CONCLUSIONS

The fast CASCIE code was developed as an analytical-numerical description tool for studying the properties of structured waveguides with an inhomogeneous structure. In this article, we presented the results of testing this code. We used this code to numerically tune the resonator chain, which is very similar to the classic 3m SLAC accelerating structure. During testing, we discovered a problem that, in our opinion, was overlooked in the development of the accelerating sections, namely the appearance of a constant phase shift when determining the dimensions of the resonator radii from the characteristics of homogeneous structures. This issue is particularly relevant in the development of short accelerating structures [41,42]

The developed CASCIE code is fast (on a laptop with 11th Gen Intel(R) Core(TM) i5-1135G7 @ 2.40GHz processor it takes less than 5 sec to compute the characteristics of the described above chain) and accurate (the relative error of electric field component calculation is less than several units of $10^{-3}$).

## ACKNOWLEDGEMENTS

The author would like to thank David Reis for his support and interest in work.

## APPENDIX 1

The simulation results presented above show that with the "homogeneous" approach an additional phase shift of the same sign between the resonators occurs in an inhomogeneous chain (linear growth of the phase deviation).

This result can be confirmed within a simple model of coupled resonators

$$Z_k e_{010}^{(k)} = e_{010}^{(k-1)} \alpha_{010}^{(k,k-1)} + e_{010}^{(k+1)} \alpha_{010}^{(k,k+1)}, \tag{13}$$

where $Z_k = \left(1 - \frac{\omega^2}{\omega_{010}^{(k)2}} - \alpha_{010}^{(k,k)} - i\frac{\omega}{\omega_{010}^{(k)}Q}\right)$, $\alpha_{010}^{(k,k)} = -v\frac{a^{(k)3} + a^{(k+1)3}}{b^{(k)2}d^{(k)}}$, $\alpha_{010}^{(k,k-1)} = v\frac{a^{(k)3}}{b^{(k)2}d^{(k)}}$, $\alpha_{010}^{(k,k+1)} = v\frac{a^{(k+1)3}}{b^{(k)2}d^{(k)}}$,

$v = \frac{2}{3\pi J_1^2(\lambda_{01})}$, $a^{(k)}$ - the hole radius between $k-1$ and $k$ resonators, $b^{(k)}$ - the radius of the $k$-th cylindrical resonator, $d$ - the resonator length, $\omega_{010}^{(k)} = \frac{c\lambda_{01}}{b^{(k)}}$, $J_0(\lambda_{01}) = 0$, $e_{010}^{(k)}$ - a complex amplitude of $E_{010}$ resonator eigen oscillation.

If a chain is homogeneous and the resonator radii fulfil the equation

$$1 - \frac{\omega^2 b^2}{c^2 \lambda_{01}^2} - v\frac{2a^3}{b^2 d} = v\frac{2a^3}{b^2 d}\cos\varphi, \tag{14}$$

the equation (13) has solution $e_{010}^{(k)} = e_{010}\exp(ik\varphi - k\gamma)$, where

$$\gamma = \frac{\omega b b^2 d}{c\lambda_{01} Q 2\sin\varphi v a^3} = \frac{\lambda_{01} d}{2Qb\beta_g} \ll 1. \tag{15}$$

The group velocity is defined as

$$\beta_g = \frac{d}{c}\frac{d\omega}{d\varphi} = v\frac{a^3 c \lambda_{01}^2}{b^2 \omega b^2}\sin\varphi. \tag{16}$$

The equation (13) is a difference equation of the second order

$$y_{k+2} + f_{1,k} y_{k+1} + f_{0,k} y_k = 0, \tag{17}$$

where

$$f_{1,k} = -2\frac{a^{(k)3}}{a^{(k+1)3}}\left(\cos\varphi_0 - i\frac{\omega b^{(k)3}d}{c\lambda_{01}Q2a^{(k)3}}\right),$$

$$f_{0,k} = \frac{a^{(k)3}}{a^{(k+1)3}}. \tag{18}$$

If the sequences $f_{0,k}$ and $f_{1,k}$ vary sufficiently slowly with $k$, the equation (17) has an approximate solution [43, 44]

$$y_k^{(1,2)} \approx \frac{1}{\left(f_{1,k}^2 - 4f_{0,k}\right)^{1/4}} \prod_{s=k_0+1}^{k} \rho_{s-1}^{(1,2)} \exp\left(\pm \sum_{s=k_0+1}^{k} \frac{f_{1,s}-f_{1,s-1}}{2\sqrt{f_{1,s}^2-4f_{0,s}}}\right), \tag{19}$$

where $\rho_k$ are the solutions of the characteristic equations

$$\rho_k^2 + f_{1,k}\rho_k + f_{0,k} = 0. \tag{20}$$

If the resonator radii fulfil the "homogeneous" equation (14)

$$1 - \frac{\omega^2 b^{(k)2}}{c^2\lambda_{01}^2} - \nu\frac{2a^{(k)3}}{b^{(k)2}d} = \nu\frac{2a^{(k)3}}{b^{(k)2}d}\cos\varphi_0, \tag{21}$$

then the solution of (20) can be written as

$$\rho_k \approx \exp\left(i\varphi^{(k)}\right)\left\{1 - \frac{\omega b^{(k+1)3}d}{2c\lambda_{01}Qa^{(k+1)3}\nu\sin\varphi_0} - \left(\frac{a^{(k+2)3}-a^{(k+1)3}}{2a^{(k+1)3}}\right)\right\}. \tag{22}$$

where $\varphi^{(k)} = \varphi_0 + \delta\varphi^{(k)}$ and

$$\delta\varphi^{(k)} = -\frac{1-\cos\varphi_0}{2\sin\varphi_0}\left(\frac{a^{(k+2)3}-a^{(k+1)3}}{a^{(k+1)3}}\right). \tag{23}$$

If coupling radii satisfy the condition

$$a^{(k+2)3} = a^{(k+1)3} - \frac{\omega b^{(k+1)3}d}{c\lambda_{01}Q\nu\sin\varphi_0}, \tag{24}$$

then from (22) it follows that the module of $\rho_k$ equals 1.

It can be shown that in our case $\varphi \approx \varphi_0$ the sum in the exponent of (19) is small (compere with [44]) and the WKB solution takes the form

$$y_k \approx \frac{1}{\sqrt{i2\sin\varphi_0}}\exp(ik\varphi_0)\prod_{s=k_0+1}^{k}\left\{1-\frac{\omega b^{(s+1)3}d}{2c\lambda_{01}Qa^{(s+1)3}\nu\sin\varphi_0}-\left(\frac{a^{(s+2)3}-a^{(s+1)3}}{2a^{(s+1)3}}\right)\right\}\exp\left(i\sum_{s=k_0+1}^{k}\delta\varphi_k\right). \tag{25}$$

We can estimate the phase deviation from the $k\varphi_0$ law

$$\sum_{s=k_0+1}^{k}\delta\varphi_k = \frac{\cos\varphi_0-1}{2\sin\varphi_0}\sum_{s=k_0+1}^{k}\left(\frac{a^{(s+2)3}-a^{(s+1)3}}{a^{(s+1)3}}\right) \approx \frac{\cos\varphi_0-1}{2\sin\varphi_0}\int_{a^{(1)3}}^{a^{(k)3}}\frac{dx}{x} =$$
$$= \frac{1-\cos\varphi_0}{2\sin\varphi_0}\ln\frac{a^{(1)3}}{a^{(k)3}} = \frac{1-\cos\varphi_0}{2\sin\varphi_0}\ln\frac{\beta_g^{(1)}}{\beta_g^{(k)}} \tag{26}$$

Therefore, if we select values of resonator radii from the "homogeneous" relation (21) we obtain additional phase shift (26) that is determined by the initial and end values of group velocities and do not depend on the number of cells. For $\beta_{g,1} = 0.02$, $\beta_{g,2} = 0.0062$ and $\varphi_0 = 2\pi/3$ we get $\sum_{s=k_0+1}^{k}\delta\varphi_k \approx 60°$. This is a smaller value than was obtained in the full simulation.